\documentclass[apjl]{emulateapj}
\usepackage{apjfonts}
\submitted{Received 2010 April 5; accepted 2010 May 21; published --}
\journalinfo{Accepted to ApJL, 2010 May 21}
\shorttitle{BACK REACTION ASSOCIATED WITH SOLAR ERUPTIONS}
\shortauthors{WANG \& LIU}

\begin{document}

\title{Observational Evidence of Back-reaction on the Solar Surface Associated with Coronal Magnetic Restructuring in Solar Eruptions}

\author{Haimin Wang and Chang Liu}

\affil{Space Weather Research Laboratory, New Jersey Institute of Technology, University Heights, Newark, NJ 07102-1982, USA;\\haimin.wang@njit.edu, chang.liu@njit.edu}

\begin{abstract}
Most models of solar eruptions assume that coronal field lines are anchored in the dense photosphere and thus the photospheric magnetic fields would not have rapid, irreversible changes associated with eruptions resulted from the coronal magnetic reconnection. Motivated by the recent work of Hudson, Fisher \& Welsch (2008) on quantitatively evaluating the back reaction due to energy release from the coronal fields, in this Letter we synthesize our previous studies and present analysis of new events about flare-related changes of photospheric magnetic fields. For the 11 X-class flares where vector magnetograms are available, we always find an increase of transverse field at the polarity inversion line (PIL) although only 4 events had measurements with 1 minute temporal resolution. We also discuss 18 events with 1 minute cadence line-of-sight magnetogram observation, which all show prominent changes of magnetic flux contained in the flaring $\delta$ spot region. Except in one case, the observed limb-ward flux increases while disk-ward flux decreases rapidly and irreversibly after flares. This observational evidence provides support, either directly or indirectly, for the theory and prediction of Hudson, Fisher \& Welsch that the photospheric magnetic fields must respond to coronal field restructuring and turn to a more horizontal state near the PIL after eruptions.
\end{abstract}


\section{INTRODUCTION}
Solar eruptions, including flares, filament eruptions, and coronal mass ejections (CMEs) have been understood as the result of magnetic reconnection in the solar corona (e.g., Kopp \& Pneuman 1976; Antiochos et al. 1999). Although surface magnetic field evolution (such as new flux emergence and shear motion) play important roles in building energy and triggering eruption, most models of flares and CMEs have the implication that photospheric magnetic fields do not have rapid, irreversible changes associated with the eruptions. The key reason behind this assumption is that the solar surface, where the coronal magnetic fields are anchored, has much higher density and gas pressure than the corona. Recently, we note the work by Hudson, Fisher \& Welsch (2008, hereafter HFW08), who quantitatively assessed the back reaction on the solar surface and interior resulting from the coronal field evolution required to release energy, and made the prediction that after flares, the photospheric magnetic fields become more horizontal. Their analysis is based on the simple principle that any change of magnetic field energy must lead to a corresponding change in magnetic pressure. This is one of the very few models that specifically predict that flares can be accompanied by rapid and irreversible changes of photospheric magnetic fields. Over a decade ago, Melrose (1997) used the concept of reconnection between two current-carrying systems to provide explanation for the enhancement of magnetic shear at the flaring magnetic polarity inversion line (PIL), which is sometimes observed (see below). Perhaps this is in the same line as the tether-cutting model for sigmoids, which was elaborated by Moore et al. (2001) and involves a two-stage reconnection processes. At the eruption onset, the near-surface reconnection between the two sigmoid elbows produces a low-lying shorter loop across the PIL and a larger twisted flux rope connecting the two far ends of the sigmoid. The second stage reconnection occurs when the large-scale loop cuts through the arcade fields causing erupting plasmoid to become CME and precipitation of electrons to form flare ribbons. If scrutinizing the magnetic topology close to the surface, one would find a permanent change of magnetic fields that conforms to the scenario of HFW08: the magnetic fields turn more horizontal near the flaring PIL due to the newly formed short loops there.

\begin{deluxetable*}{lccccc}
\tablewidth{0pt}
\tablecaption{Events with Evidences of Enhancement of Transverse Magnetic Field ($B_t$) at the PIL}
\tablehead{\colhead{Event Date}  & \colhead{{\it GOES} Level} & \colhead{Data Source} & \colhead{Data Cadence} & \colhead{Main Findings} & \colhead{References}}
\startdata
(1) 1990 Aug 27   & X3.0  & BBSO  & 10 min & magnetic shear increase & 1 \\
(2) 1991 Mar 22   & X9.0  & BBSO  & 10 min & magnetic shear increase & 2 \\
(3) 1991 Jun 9   & X10.0  & BBSO/HSO  & 5 hr & magnetic shear increase & 2 \\
(4) 2000 Jul 14   & X5.7  & HSO  & 3 hr & $B_t$ and electric current increase & 3 \\
(5) 2001 Aug 25   & X5.3  & BBSO & 1 min & $B_t$ and magnetic shear increase & 4 \\
(6) 2001 Oct 19   & X1.6  & BBSO  & 1 min & $B_t$ and magnetic shear increase & 4 \\
(7) 2002 Jul 26   & M8.7  & BBSO & 1 min & $B_t$ increase & 5 \\
(8) 2003 Oct 29   & X10  & MSFC & 3 hr & $B_t$ and magnetic shear increase & 6 \\
(9) 2005 Jan 15   & X2.6  & BBSO & 1 min & $B_t$ and inclination angle increase & 7 \\
(10) 2005 Sep 13   & X1.5  & BBSO & 1 min & $B_t$ increase & 8 \\
(11) 2006 Dec 13   & X3.4  & Hinode  & 8 hr & $B_t$ and magnetic shear increase & 9
\enddata
\tablecomments{References: (1) Wang 1992; (2) Wang et al. 1994; (3) Wang et al. 2005; (4) Wang et al. 2002; (5) Wang et al. 2004a; (6) Liu et al. 2005; (7) Li 2010; (8) Wang et al. 2007; (9) Jing et al. 2008}
\end{deluxetable*}

On the observational side, earlier studies were inconclusive on the flare-related changes of photospheric magnetic field topology. Wang (1992) and Wang et al. (1994) showed impulsive changes of vector fields after flares including some unexpected patterns such as increase of magnetic shear along the PIL, while mixed results were also reported (Ambastha et al. 1993; Hagyard et al. 1999; Chen et al. 1994, Li et al. 2000a, 2000b). It is not until recently that rapid and permanent changes of photospheric magnetic fields, mainly the line-of-sight component, are observed to be consistently appear in major flares and considered as indicative of flare energy release (Kosovichev \& Zharkova 2001; Sudol \& Harvey 2005). In particular, a number of papers of Big Bear Solar Observatory (BBSO)/New Jersey Institute of Technology group have been devoted to the finding of sudden unbalanced magnetic flux change (Spirock et al. 2002; Wang et al. 2002; Yurchyshyn et al. 2004; Wang et al. 2004a; Wang 2006) and a new phenomenon of sunspot white-light structure change (Wang et al. 2004b; Deng et al. 2005; Liu et al. 2005; Chen et al. 2006) associated with flares. By evaluating the mean relative motions between two magnetic polarities of five flaring $\delta$ spots, Wang (2006) revealed a sudden release of the overall magnetic shear and found a correspondence between converging/diverging motion and increase/decrease of magnetic gradient at the PIL, which suggests magnetic reconnection at or close to the photosphere. Liu et al. (2005) discussed outer penumbral decay and central penumbral darkening of $\delta$ spot seen in seven X-class flares, and proposed a reconnection picture where the active region field collapses inward after flares as signified by HFW08.

Although studying the change of magnetic field has promise to advance the understanding of flare energy release process, it is notable that the long-accumulated observational evidence have not yet converged regarding how the observed changes of photospheric magnetic fields could be understood in the context of models of coronal magnetic field restructuring. The objective of this Letter is hence to examine the observations in a systematic fashion and compare them quantitatively with the prediction of HFW08. Meanwhile, several conflicting concepts in our earlier papers will be reconciled with new physical understanding. We anticipate that these will guide more efficient studies to analyze vector magnetograph data of {\it Helioseismology and Magnetic Investigation} ({\it HMI}) on board {\it Solar Dynamic Observatory} ({\it SDO}) in the near future.

\begin{figure}
\epsscale{1.15}
\plotone{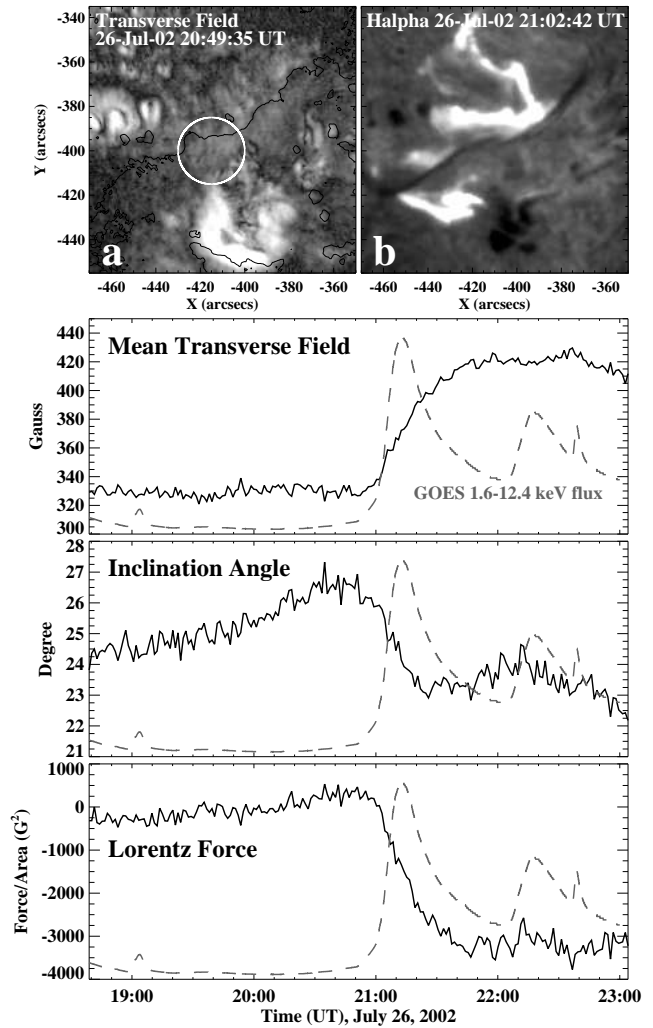}
\caption{Time profiles of transverse field, inclination angle, and Lorentz force per unit area within a white circled region (in $a$) at the PIL (black line in $a$) for the 2002 July 26 M8.7 flare (see an H$\alpha$ image in $b$), calculated using BBSO vector magnetograms. In {\it GOES} 10 soft X-ray flux (dashed line), the flare started at 20:51~UT, peaked at 21:12~UT, and ended at 21:29~UT.}
\end{figure}

\begin{deluxetable*}{lcccccc}
\tablewidth{0pt}
\tabletypesize{\scriptsize}
\tablecaption{Events with Evidences of Rapid Changes of Active Region Line-of-Sight Magnetic Fluxes \label{table 2} }
\tablehead{\colhead{Event}  & \colhead{NOAA AR} & \colhead{Location (deg)} & \colhead{{\it GOES} Level} & \colhead{Limbward Flux (10$^{20}$ Mx)} & \colhead{Diskward Flux (10$^{20}$ Mx)} & \colhead{References}}
\startdata
(1) 1991 Mar 22 & 06555  & E20 S23  & X9.0  & $+1.0$ & 0 & 1 \\
(2) 2001 Apr 2 & 09393  & W64 N19  & X20.0  & $+6.0$ & $-1.5$ & 1, 4 \\
(3) 2001 Apr 6 & 09415  & E30 S20  & X5.6  & $+4.0$ & 0.0 & 1 \\
(4) 2001 Apr 9 & 09415  & W04 S21  & M8.0  & $+2.7$ & 0 & 2 \\
(5) 2001 Aug 25 & 09591  & E34 S17  & X5.3  & $+1.8$ & $-0.8$ & 1 \\
(6) 2001 Sep 24 & 09632  & E18 S18  & X2.6  & $+10$ & $-10$ & 5 \\
(7) 2001 Oct 19 & 09661  & W29 N15  & X1.6  & $+3.0$ & $-0.4$ & 1 \\
(8) 2001 Oct 22 & 09672  & E16 S18	&  X1.2	& $+11$	& $-2$ & 1\\
(9) 2001 Oct 25 & 09672 & W26 S19 & X1.3 &	$-3$ & $+5$ & 5\\
(10) 2002 Jul 23 & 10039 & E54 S12 &	X4.8 &	$+0.15$	& $-0.8$ & 3\\
(11) 2003 Mar 18 & 10314 & W53 S16	& X1.5	& $+5.0$ & 0 &5 \\
(12) 2003 Oct 19 & 10484 & E54 N05	& X1.1	& $+7$ & $-2$ & 1\\
(13) 2003 Oct 26 & 10484 & N04 W41 & X1.2 & 0 & 0  & 5\\
(14) 2003 Oct 28 & 10486 & E04 S16 & X17.2 & $+0.75$ & $-0.4$ & 2\\
(15) 2003 Oct 29 & 10486 & W10 S17 &	X10.0	& $+0.65$ & $-0.33$ & 2\\
(16) 2004 Jul 16 & 10649 & S10 E26 & X3.6 & $+2.5$ & $-10.0$ & 5\\
(17) 2004 Nov 7 & 10696 & N09 W08 & X2.0 & $+5.0$ & $-3.0$ &5 \\
(18) 2005 Sep 13 & 10808 & E17 S11 &	X1.5 & $+0.75$ & $-1.2$ & 2\
\enddata
\tablecomments{References: (1) Wang et al. 2002; (2) Wang 2006; (3) Yurchyshyn et al. 2004; (4) Spirock et al. 2002; (5) This paper}
\end{deluxetable*}

\section{OBSERVATIONS AND RESULTS}
The most straightforward way to determine changes of vector fields associated with flares is to monitor the time sequence of vector magnetograms. In order to detect any rapid and subtle variation, however, magnetogram observations with high cadence (a few minutes), high resolution (1\arcsec), and high polarization accuracy are required, and it is understood that {\it HMI}/{\it SDO} will be able to provide unprecedented data with these characteristics. Yet in the past two decades there are some vector magnetograms available that can tackle this topic with certain limitations. In Table 1, we summarize studies of changes of vector fields in previous publications. The data source was mostly the vector magnetograph of BBSO, which typically covers a field of view of $\sim$300\arcsec~$\times$~300\arcsec\ with an often cadence for a complete set of Stokes images of 1 minute. It has been noted that (1) the change of the measured field structure is subject to seeing variation; and (2) the system utilizes one position in the Ca~{\sc i}~6103~\AA\ spectrum line thus the measured magnetic fields are saturated in regions with strong field strength. Nevertheless, it is almost impossible that the seeing fluctuation reproduces an identical trend of flare-related changes in multiple events. For almost all the events, we concern ourselves with the magnetic fields at the PIL where the weak field approximation is usually acceptable. Other ground-based magnetographs used include the filtergraph-based systems of Huairou Solar Observing station (HSO) of Beijing Astronomical Observatory and the Marshal Space Flight Center (MSFC), which have lower data cadence and bear similar restrictions as BBSO. For the 2006 December 13 event, vector magnetograms with highest polarization accuracy and resolution and taken under the seeing-free condition were obtained with the spectral polarimeter (SP) of the {\it Solar Optical Telescope} onboard {\it Hinode}, but the cadence of SP data is usually a few hours.

With these cautions in mind, it is nonetheless obvious that all the results thus far point to the conclusion that transverse magnetic field strength increases at the PIL after major flares (namely, the fields there turn more horizontal), which provides direct and strong observational support for the theory of HFW08. Please note that (1) this kind of irreversible change of field strength at the PIL is not the magnetic transient that is caused by flare emissions and is hence co-spatial with flare ribbons or kernels (Qiu \& Gary 2003; Patterson \& Zirin 1981); and (2) our earlier studies were concentrated on the magnetic shear and it has been demonstrated that increases of magnetic shear and transverse field strength are interrelated (e.g. Wang et al. 2002). Following HFW08, we further quantify the change of Lorentz force per unit area in the vertical direction using the formula:

\begin{equation}
\delta f_z=(B_z\delta B_z-B_x\delta B_x-B_y\delta B_y)/4\pi \ ,
\end{equation}

\noindent and present an example of our re-analysis of the vector field observations associated with the 2002 July 26 M8.7 flare in Figure 1. For the compact region at the flaring PIL (the circled area in Fig.~1$a$), the results unambiguously show the following. First, the mean transverse field strength increases $\sim$90~G in about one hour ensuing from the rapid rising of the flare soft X-ray emission at 20:51~UT. This change-over time of transverse field strength between the pre- and post-flare states is longer than that of the most other events as well as that of the line-of-sight field reported in Sudol \& Harvey (2005). Second, the inclination angle decreases $\sim$3$^{\circ}$ accordingly, which indicates that magnetic field lines there turn to a more horizontal direction as predicted by HFW08. Third, the change of the Lorentz force per unit area is $\sim -5000$~G$^2$, integrated which over the analyzed area of $\sim$3.2~$\times$~10$^{18}$~cm$^2$ yields a downward net Lorentz force in the order of $1.6$~$\times$~10$^{22}$~dynes, comparable to what is expected by HFW08. Similar results have been found for other flares listed in Table 1.

\begin{figure}
\epsscale{1.15}
\plotone{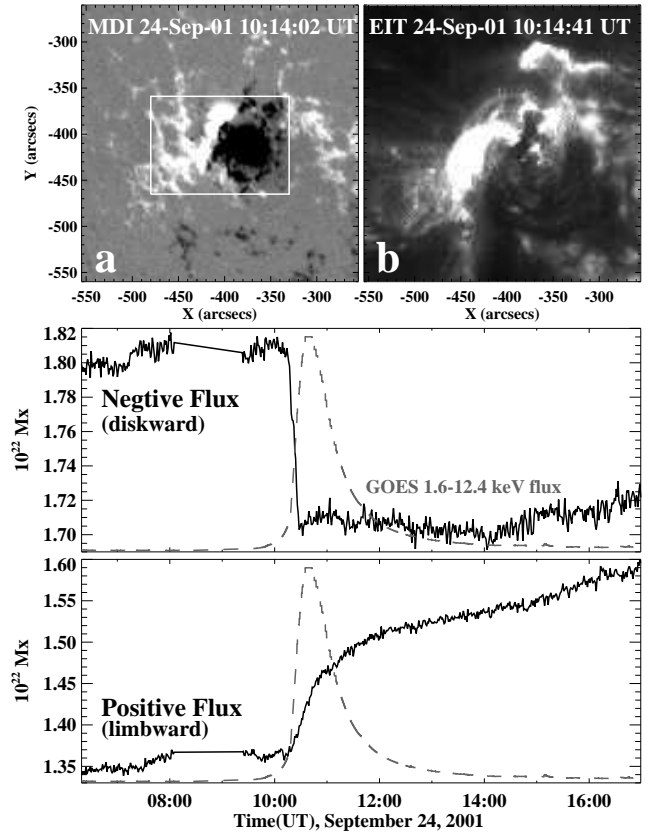}
\caption{Time profiles of negative and positive MDI line-of-sight magnetic fields within a boxed region (in $a$) covering the entire $\delta$ spot for the 2001 September 24 X2.6 flare, seen in an EUV Imaging Telescope (EIT) image ($b$). In {\it GOES} 10 soft X-ray flux (dashed line), the flare started at 09:32~UT, peaked at 10:38~UT, and ended at 11:09~UT.}
\end{figure}

\begin{figure}
\epsscale{0.8}
\plotone{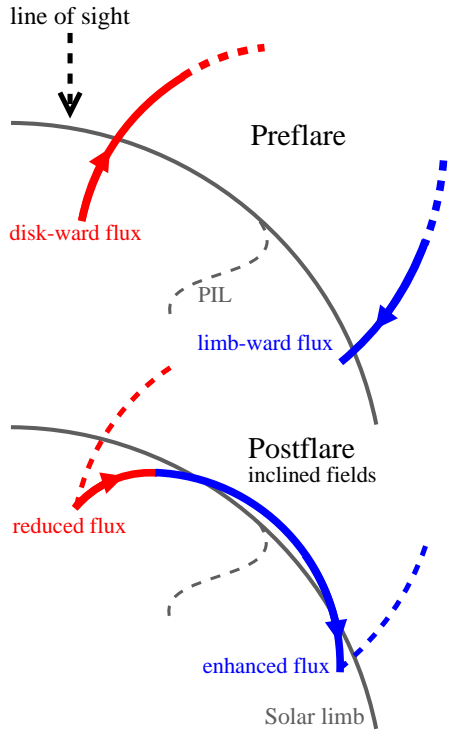}
\caption{A conceptual cartoon to demonstrate the apparent changes of line-of-sight magnetic fields when the field lines turn more horizontal with respect to the solar surface. The limbward flux would increase while diskward flux would decrease.}
\end{figure}

Considering that vector magnetograms covering major flares with sufficient cadence and quality are rare and that reliable detection of rapid changes in the observed magnetic signals would ideally require stable observing conditions, we use the line-of-sight magnetograms ($\sim$2\arcsec\ pixel$^{-1}$) measured with the Michelson Doppler Imager (MDI) onboard the {\it Solar and Heliospheric Observatory} to add more events in the present study, taking advantage of its long-sequence and seeing-free data set. In retrospect, Cameron and Sammis (1999) was the first to introduce the concept of using line-of-sight magnetograms close to limbs as a probe for the orientation of transverse magnetic fields, which was elaborated further by Wang (2006). We surveyed all the X-class flares satisfying our event selection criteria: (1) MDI magnetograms with 1 minute cadence are available spanning pre- and post-flare states for at least two hours each; (2) the source active region is within 65$^{\circ}$ from the disk center; and (3) Wang et al. (2000) concluded that there might exist two kinds of two-ribbon flares, one of which has small initial ribbon separation and occurs at the $\delta$ spot that possesses a well-defined PIL, and the other of which has wide initial separation of ribbons and is not associated with a clear and continuous PIL. We only chose the first kind of flares so that the flaring magnetic fields can be coherently determined and assessed. With these criteria, a total of 18 events are identified and are summarized in Table 2, a few of which are newly analyzed in this paper. Except for one case (2001 October 25 X1.3 flare), all events exhibit an increase of limbward flux and a decrease of diskward flux of active region magnetic fields after flares with an order of magnitude of 10$^{20}$~Mx (no obvious changes can be detected for one of the two polarities in some flares). There is no doubt that all the changes are closely related with the flare occurrence and are substantially above the noise level with the largest percentage of change of $\sim$30\%. As an instance, Figure 2 (lower panels) shows the line-of-sight magnetic field changes associated with the 2001 September 24 X2.6 flare (see Fig.~2$b$) at NOAA AR 09632, which are calculated for the boxed region of the entire $\delta$ spot (see Fig.~2$a$). It can be clearly seen that the limbward/diskward (positive/negative) fluxes increase/decrease for a similar amount of $\sim$10$^{21}$~Mx right after the flare impulsive phase. The change-over time for positive and negative fluxes is $\sim$40 and 10 minutes, respectively. As a matter of fact, these behaviors of line-of-sight fields also imply that the active region magnetic fields become more horizontal after flares, which was indicated by Wang (2006) based on limited sample events. Here we further demonstrate this linkage in Figure 3. It shows that if the source active region is not located at the disk center, the measured apparent line-of-sight fields would undergo the aforementioned unbalanced changes due to projection effect, when the field lines collapse inward, that is, turn to a more horizontal state. Therefore, the theory of HFW08 is also indirectly substantiated by our results of line-of-sight field measurement.

It is worth mentioning that one can estimate the order of magnitude of the change of Lorentz force based on line-of-sight fields only using $\delta F_z=(B_z\delta \phi_z)/4\pi$, where $F_z$ is the integrated force and $\delta \phi_z$ is the integrated change of flux. With $B_z \sim 1000$~G and $\delta \phi_z \sim 10^{20}$~Mx, $F_z$ is about 10$^{22}$ dynes similar to what is derived before using vector data.

\section{SUMMARY AND DISCUSSION}
Synthesizing the research of flare-related rapid and irreversible changes in both vector and line-of-sight magnetic fields, we have revealed that photospheric fields apparently respond to the back reaction of coronal fields due to flare energy release, in a pattern that magnetic fields near the PIL become more horizontal, which strongly evidences the conjecture of HFW08. The change-over time lies between $\sim$10 minutes to 1 hour, and all the changes are co-temporal with the flare initiation. These should provide insight into the upcoming {\it HMI} data analysis towards a more complete physical understanding of the impacts of flares on the solar atmosphere following coronal field restructuring. We discuss a few points related to observations and modeling work.

1. It is obviously reasonable that larger flares tend to produce more prominent field structure change (e.g., Chen et al. 2006). Using unprecedented {\it HMI} measurements, we anticipate finding a lower threshold of flare magnitude at which the magnetic field changes are still detectable.

2. A related aspect is that Hudson (2000) postulated that the energy conversion process during coronal transients would involve a magnetic implosion, which signifies that the flare ribbons and loops may contract initially before expansion. Although multi-wavelength signatures of such contraction have been observed recently (e.g., Liu \& Wang 2009 and references therein), it is still unclear how to link it to the changing of field line orientation from more vertical to more horizontal configuration.

3. As our previous studies have shown, it is most likely that the strengthening of transverse fields near the PIL has a causal relationship to the enhancement of central penumbral structure of $\delta$ spots (e.g., Wang et al. 2004b; Liu et al. 2005). However, the penumbral decay in the outer boundary region is not reflected in the model of HFW08. We speculate that it is due to the peripheral field lines changing to more vertical state when the central region pressure is released after flares. That is to say, the surrounding fields might be subsequently pushed inwards to fill the void.

4. HFW08 introduced the concept of a ``jerk'' (i.e., downward push in the vertical direction due to the change of Lorentz force), which may account for the launch of seismic waves seen in some impulsive flares (Kosovichev \& Zharkova 1998). This connection has other supporting evidence (e.g. Mart\'inez-Oliveros \& Donea 2009). A pertinent phenomenon is that a sunspot can have rapid flare-induced motion along the solar surface, which seems to be sufficiently driven by the changes of the horizontal components of Lorentz force (Anwar et al. 1994). Liu et al. (2010, in preparation) presented five new events in an effort to associate the kinetic energy of sunspot lateral motion with that of seismic waves, while we must note that the detailed mechanism involved has never been researched before. In addition, although the events with seismic waves always have associated rapid changes of surface magnetic fields, it seems that this conclusion cannot be reversed as suggested by several cases under study. The new research area will shed more light on understanding the linkage between the sub-surface/surface magnetic activities and coronal eruptions.

\acknowledgments
We benefited significantly from the discussions with Drs. Hugh Hudson and Brian Welsch that motivated this study. We also thank the referee
for helpful comments. This work is supported by NSF grants AGS-0839216, AGS-0819662 and AGS-0849453;  NASA grants NNX08-AQ90G and NNX08-AJ23G.


\begin{references}
\reference { } {Ambastha A., Hagyard, M. J., \& West E. A., 1993, Sol.
Phys., 148, 277}

\reference { } {Anwar, B., Acton, L. W., Hudson, H. S., Makita, M., McClymont, A. N. \& Tsuneta, S., 1993, Sol. Phys., 147, 287}

\reference { } {Antiochos, S. K., DeVore, C. R., \& Klimchuk, J. A., 1999,
Ap. J., 510, 485}

\reference { } {Cameron, R. \& Sammis, I., 1999, Ap.J. Letters, 525, L61}

\reference {} {Chen, J. Wang, H., Zirin, H. \& Ai, G., 1994, Sol.
Phys., 154, 261}

\reference {} {Chen, W., Liu, C. \& Wang, H., 2007, ChJAA, 7, 733}

\reference {} {Deng, N., Liu, C., Yang, G., Wang, H., \& Denker C.,
2005, Ap.J., 623, 1195}

\reference { } {Hagyard et al., 1999, Sol. Phys., 184, 133}

\reference {}  {Hudson, H. S., 2000, Ap.J. Letters, 531, L75}

\reference {} {Hudson, H. S., Fisher, G. H. \& Welsch, B.T., 2008, ASP Conference Series, 383, 221}
\reference {} {Jing, J., Wiegelmann, T., Suematsu, Y., Kubo, M. \& Wang, H., 2008, Ap.J. Letters, 676, L81}

\reference { } {Kopp, R. A. \& Pneuman, G. W. 1976, Sol. Phys., 50, 85}

\reference {}  {Kosovichev, A. G. \& Zharkova, V. V., 1998, Nature, 393, 317}

\reference { } {Kosovichev, A. G. \& Zharkova, V. V. 2001,
Ap.J. Letters, 550, L105}

\reference {} {Li, H., Sakurai, T., Ichimoto, K. \& UeNo, S., 2000a, PASJ, 52,
465}

\reference {} {Li, H., Sakurai, T., Ichimoto, K. \& UeNo, S., 2000b, PASJ, 52, 483}
\reference {} {Li, Y., 2010, Ph.D. Thesis Proposal, New Jersey Institute of Technology}

\reference {} {Liu, C., Deng, N., Liu, Y., Falconer, D., Goode,
P.~R., Denker, C., \& Wang, H., 2005, Ap.J., 622, 722}

\reference {}  {Liu, R. \& Wang, H., 2009, Ap.J. Letters, 703, L23}

\reference {}  {Mart\'inez-Oliveros, J.C. \& Donea, A.C., 2009, MNRAC, 395,39}


\reference {} {Melrose, D.B., 1997, Ap.J., 486, 521}

\reference { } {Moore, R.L., Sterling, A.C., Hudson, H.S., \& Lemen, J., 2001,
Ap.J., 552, 833}

\reference {} {Patterson, A. \& Zirin, H., 1981, Ap.J. Letters, 243, L99}

\reference { } {Qiu, J. \& Gary, D.E., 2003, Ap.J., 599, 615}

\reference { } {Spirock, T. J., Yurchyshyn, V. B., \& Wang, H., 2002, Ap.
J., 572, 1072}

\reference {}  {Sudol, J.J. \& Harvey, J.W., 2005, Ap.J., 635, 647}

\reference {}  {Wang, H., 1992, Sol. Phys., 140, 85}

\reference {}  {Wang, H., 2006, Ap.J., 649, 490}

\reference { } {Wang, H., Ewell, M.W., Zirin, H. \& Ai, G.
1994, Ap.J., 424, 436}

\reference {} {Wang, H., Goode, P. R., Denker, C., Yang, G., Yurchyshyn, V., Nitta, N., Gurman, J. B., St. Cyr, C. \& Kosovichev, A. G.	, 2000, Ap.J., 536, 971}

\reference {} {Wang, H., Liu, C., Jing, J., Yurchyshyn, V., 2007, Ap.J., 671, 973}

\reference {} { Wang, H., Liu, C., Qiu, J., Deng, N., Goode, P. R.,
\& Denker, C., 2004b, Ap.J. Letters, 601, L195}

\reference { } {Wang, H., Liu, C., Zhang, H. \& Deng, Y, 2005, Ap.J.,
627, 1031}
\reference {} {Wang, H., Qiu, J., Jing, J., Spirock, T. J. \&
Yurchyshyn, V., 2004a,  Ap.J., 605, 931}

\reference {} {Wang, H., Spirock, T. J., Qiu, J., Ji, H.,
Yurchyshyn, V., Moon, Y.J., Denker, C. \& Goode, P. R., 2002,
Ap.J., 576, 497}

\reference { } {Yurchyshyn, V. B., Wang, H., Abramenko, V., Spirock,
T. J., \& Krucker, S., 2004,  Ap.J., 605, 546}

\end{references}
\end{document}